\documentclass[12pt]{article}
\usepackage{epsfig}
\usepackage{psfig}
\usepackage{graphicx}
\usepackage{amsmath}
\usepackage{amsfonts}
\usepackage{amssymb}

\begin{document}

\date{}
\title{\textbf{Dynamical Casimir effect with cylindrical waveguides}}
\author{ P. A. Maia Neto
\\{\small \textit{Instituto de F{\'\i}sica, UFRJ, Caixa Postal
68528, Rio de Janeiro RJ, 21945-970, Brasil}}}
\date{\today}
\maketitle
\begin{abstract}
I consider the quantum electromagnetic field in a coaxial cylindrical waveguide, such that the outer cylindrical surface
has a time-dependent radius. The field propagates parallel to the axis, inside the annular region between the 
two cylindrical surfaces. 
When the mechanical frequency and the thickness of the annular region  are small enough, only 
Transverse Electromagnetic (TEM) photons may be 
generated by the dynamical Casimir effect. The photon emission rate is calculated
in this regime, and compared with the case of parallel plates in the limit of very short distances between the two cylindrical surfaces. The proximity force 
approximation holds for the transition matrix elements in this limit, but the emission rate 
scales quadratically with the mechanical frequency, as opposed to the cubic dependence for parallel plates.  
\end{abstract}

The reflection of vacuum fluctuations by an oscillating mirror generates frequency sidebands. 
This frequency modulation mixes up positive and negative field frequencies below the mechanical 
frequency. Because of the association between positive (negative) frequencies and annihilation
(creation) operators, this mixing leads to the creation of real low-frequency photon pairs out of the vacuum field state. 
The very small orders
of magnitude involved in this ``dynamical Casimir" effect
have so far ruled out its experimental verification. For a single plate moving in vacuum, 
the photon emission rate is of the order of a few microwave photons per day
even for mechanical frequencies as high as 10 GHz \cite{PAMN1}.

Much higher orders of magnitude are obtained if the mechanical frequency is tuned into 
parametric resonance with a microwave high-$Q$ closed cavity \cite{closed}.  
Another possibility is to consider the dynamical counterpart of the 
original configuration proposed by Casimir~\cite{Casimir}, with two parallel 
plates moving along the direction perpendicular to the plates ($z$ direction)~\cite{two-plates}. 
Assume that one of the plates oscillates with frequency $\omega_0$: $\delta z_1(t)=\delta z_0\cos\omega_0 t$, and the 
second plate is at rest at $z_2=a.$ Since this `cavity' is laterally open, the spectrum of field modes is 
continuous, and there is no re-cycling of photons (except in the very particular
case of propagation along the $z$ direction). 
As a consequence, one may employ a perturbative approach, 
which is implemented either by considering the frequency sidebands discussed above, or alternatively by
using standard Hamiltonian perturbation theory. 
Since there is a continuous of final states, from Fermi golden rule
the average photon number is proportional to  time, allowing one 
to define the photon production rate as the physical quantity of interest.

For short values of the
average distance $a$, $\omega_0 a /(\pi c) < 1,$ the photon production rate is given by~\cite{two-plates}
\begin{equation}\label{two-plates}
\frac{dN}{dt} = \frac{1}{64} \frac{A \omega_0^3 \,\delta z_0^2}{c^2 a^2},
\end{equation}
where $A$ is the area of the plates. 
As compared to the single-plate configuration, the emission rate 
 is enhanced by the factor $1/(\omega_0 a/c)^2$
for short distances.
If we take $\omega_0/(2\pi) = 10 {\rm GHz},$   
$v_0=\omega_0 \delta z_0 = 10^{-7} c,$ and $A \omega_0^2/c^2 =(2\pi)^2$ [$A=(3{\rm cm})^2$] we 
find 
\begin{equation}\label{order1}
\frac{dN}{dt} = 8.8\times 10^3 \,{\rm sec}^{-1}
\end{equation}
for $a=1\mu{\rm m}.$ 
In this regime, all photons have 
frequency  $\omega_0/2,$ and Transverse Magnetic (TM) polarization (magnetic field perpendicular to the 
$z$ axis). 
The electric field is parallel to the 
$z$ axis and constant along this direction. In waveguide language, this corresponds to TEM polarization, 
since both electric and magnetic fields are perpendicular to the direction of 
propagation, which is parallel to the $xy$ plane.
The `twin' photons in a given pair propagate along opposite directions, 
and the total production rate is uniformly distributed over the $2\pi$ angle in the $xy$ plane. 
In this sense, the parallel-plates configuration is a not an efficient waveguide, because 
all directions in the $xy$ plane are equivalent by symmetry.   

In this paper, I consider the dynamical Casimir effect in a coaxial cylindrical 
waveguide. 
As compared with 
the parallel-plates setup, it has the advantage of concentrating the 
emission rate over the two opposite directions along the $z$ axis.    
The inner cylinder has radius $b$, whereas the outer cylindrical surface has a time-dependent
radius $b+a+\delta \rho(t)$, with
\begin{equation} \label{deltarho}
\delta \rho(t)=\delta \rho_0 \cos\omega_0 t.
\end{equation}
The field propagates in the annular region of thickness
$a+\delta \rho(t)$ between the two perfectly-reflecting cylindrical surfaces. 

We assume that $\omega_0$ is below the frequency cutoff for
propagation of TE and TM waveguide modes.
 Since the emitted photons 
have frequencies below $\omega_0$, they necessarily correspond to TEM modes. Thus, from now on
we only analyze these modes. We take periodic boundary conditions over the axial length $L$, which is to be
identified with the (very long) length of the waveguide. Then, the electric field is given by the Fourier series
\[
{\bf E}^{\rm\scriptscriptstyle TEM}(\rho,\phi,z,t) = \sum_{n=-\infty}^{\infty} 
{\bf E}^{\rm\scriptscriptstyle TEM}_n(\rho,\phi,t) \exp( in\pi z/L)
\] 
with
\begin{equation}\label{En}
{\bf E}^{\rm\scriptscriptstyle TEM}_n(\rho,\phi,t) =
 i\sqrt{\frac{\hbar \omega_n}{4\pi\epsilon_0 L \log(1+a/b)}} \,
\frac{a_n e^{-i\omega_n t} {\hat \rho}}{\rho} + {\rm H.c.},
\end{equation}
$[a_n,a_{n'}^{\dagger}]=\delta_{n,n'},$
and  $\omega_n= |n| \pi c/L.$ 
The magnetic field is given by
${\bf B}^{\rm\scriptscriptstyle TEM}_n= {\rm sgn}(n) {\hat z}\times {\bf E}^{\rm\scriptscriptstyle TEM}_n/c.$

The effect of the motion is modeled by the perturbation Hamiltonian
representing the negative of the work done on the expanding/contracting cylindrical surface:
\begin{equation}\label{deltaV}
\delta V(t) = -(a+b) \int_0^{2\pi} d\phi \int_0^L dz 
\lim_{\epsilon\rightarrow 0^+}T_{\rho\rho}(a+b-\epsilon,\phi,z) \,\delta\rho(t), 
\end{equation}
where 
\begin{equation}\label{Trhorho}
T_{\rho\rho} = \frac{\epsilon_0}{2}\left(E_{\rho}^2-B_{\phi}^2 \right) 
\end{equation}
is the suitable component of the Maxwell stress tensor. 
Being quadratic in the bosonic field operators, it leads to the generation of photon pairs 
within first-order time-dependent perturbation theory. We assume 
the field to be initially in the vacuum state, $|\psi\rangle_0=|0\rangle,$ and 
write the field state at time $\Delta t$ as 
\begin{equation}\label{psi}
|\psi \rangle_{\Delta t} = b(\Delta t) |0\rangle 
+\sum_{\{n_1,n_2\}} c_{\{n_1 n_2\}}(\Delta t)|\{n_1 {\scriptstyle\rm  TEM};\,n_2 {\scriptstyle\rm  TEM}\}\rangle,
\end{equation}
where the sum is over all pairs of integers $\{n_1,n_2\}$ regardless of the ordering, so that 
each two-photon state appears only once in (\ref{psi}). 
The two-photon amplitudes are calculated with the help of first-order perturbation theory:
\begin{equation}\label{c}
c_{\{n_1 n_2\}}(\Delta t) = -\frac{i}{\hbar}\int_{0}^{\Delta t}
\langle\{ n_1{\scriptstyle\rm  TEM};\,n_2 {\scriptstyle\rm  TEM}\}|\delta V(t')|0\rangle 
\exp[i(\omega_{n_1}+\omega_{n_2})t'] \,dt'.
\end{equation}

The matrix elements appearing in (\ref{c}) are computed from (\ref{En}) and (\ref{Trhorho})
after replacing the integral over $z$ in (\ref{deltaV}) by a sum over Fourier components. 
The electric and magnetic contributions are equal, yielding 
\begin{equation}\label{matrix-element}
\langle\{ n_1 {\scriptstyle\rm TEM}; n_2 {\scriptstyle\rm TEM}\}|\delta V(t)|0\rangle 
= \frac{\hbar\omega_{n_1}\delta\rho(t)}{(a+b)\log(1+a/b)} \delta_{n_1,-n_2}.
\end{equation}
At this point, it is illustrative to compare this result with the case of two parallel plates discussed 
in the beginning.
The matrix elements of the field operator representing the force on the 
left-hand plate are given by 
Eq.~(47) of Ref.~\cite{two-plates} (correcting a misprint):
\begin{equation}\label{plates}
\langle \{ {\bf n}_1 \ell_1 {\scriptstyle\rm TM}, {\bf n}_2 \ell_2 {\scriptstyle\rm TM} \}|F|0\rangle = \frac{\hbar}{a} 
\frac{ c^2(k_{\|}^{{n}_1})^2 + 
\omega_{n_1}^{\ell_1}\omega_{n_1}^{\ell_2}}
{\sqrt{(1+\delta_{\ell_1 0})(1+\delta_{\ell_2 0})\omega_{n_1}^{\ell_1}\omega_{n_1}^{\ell_2}}}
 \delta_{{\bf n}_1,-{\bf n}_2}.
\end{equation}
The TM-polarized two-photon states in Eq.~(\ref{plates}) 
are characterized  by two sets of integer numbers $n_x,$ $n_y$ and $\ell,$ 
where ${\bf n}=n_x{\hat x}+n_y{\hat y}$   defines the component of the wavevector parallel to the plates:
${\bf k}_{\|}^{\bf n}={\bf n}\pi c/\sqrt{A};$ and $\ell$ defines the wavelength of the standing wave along the 
$z$ direction. The corresponding frequency 
is given by $\omega_{n}^{\ell}=c\sqrt{({k}_{\|}^{n})^2+(\ell \pi/a)^2}.$

As discussed above, at low mechanical frequencies and short distances, only 
 TM modes with $\ell=0$ are excited.   The corresponding 
matrix elements of the perturbation Hamiltonian $\delta V(t) = -F \delta z_1(t)$
are obtained from 
Eq.~(\ref{plates}) and written in terms of the displacement $\delta\rho=-\delta z_1$ representing 
an increase of the separation as in the problem of the coaxial waveguide:
\begin{equation}\label{plates2}
\langle\{ n_1\,0\, {\scriptstyle\rm TM}; n_2\, 0\, {\scriptstyle\rm TM}\}|\delta V(t)|0\rangle 
= \frac{\hbar\omega_{n_1}^0\delta\rho(t)}{a} \delta_{n_1,-n_2}.
\end{equation}
This coincides with the result for the matrix elements in the cylindrical waveguide 
as given by (\ref{matrix-element}) in the limit $a\ll b.$ Thus, a variant of the 
proximity force approximation applies for the transition matrix elements of the force/stress 
tensor operators, allowing for a direct comparison between the coaxial waveguide and parallel
plates configurations. 

On the other hand, the emission rates  themselves are very different, because for the parallel  plates
there are two unconstrained spatial dimensions, and only one for the cylindrical waveguide. 
The emission rate in this second configuration is derived from the
combination of the  matrix elements given by (\ref{matrix-element}) with the 
one-dimensional free-space density of modes, rather than the two-dimensional 
density for the parallel plates. 
When replacing (\ref{deltarho}) into (\ref{matrix-element}), we neglect the counter-rotating term and derive the 
probability of pair creation from (\ref{c}):
\begin{equation}\label{prob}
|c_{\{n_1 n_2\}}(\Delta t)|^2 = \left(\frac{ \omega_{n_1} \delta\rho_0}{(a+b)\log(1+a/b)}\right)^2
\frac{\sin^2(\Delta\omega\, \Delta t/2)}
{\Delta\omega^2}\,\delta_{n_1,-n_2},
\end{equation}
with 
$$\Delta\omega = \omega_{n_1}+\omega_{n_2}-\omega_0=2\omega_{n_1}-\omega_0.$$

The average
photon number at time $\Delta t$ is given by 
\begin{equation}
\Delta N = \sum_n |c_{n, -n}(\Delta t)|^2.
\end{equation}
Then we replace $\sum_n\rightarrow L\int dk/(2\pi),$
and calculate the resulting integral in the limit $\Delta t\rightarrow\infty,$ so that 
the function $\sin^2(\Delta\omega\, \Delta t/2)/\Delta\omega^2$ is sharply peaked around $\omega_n=\omega_0/2.$ 
Hence the twin photons of a given pair share the same frequency $\omega_0/2,$ and propagate along 
opposite directions along the $z$ axis because  
$n_1=-n_2$ in Eqs.~(\ref{matrix-element}) and (\ref{prob}). 
The resulting $\Delta N$ is proportional to $\Delta t,$ allowing for the 
definition of the emission rate $dN/dt=\Delta N/\Delta t,$ with
\begin{equation}
\frac{dN}{dt} = \frac{1}{16} \frac{L \omega_0^2 \,\delta\rho_0^2}{c\,(a+b)^2[\log(1+a/b)]^2}
\end{equation}
The emission rate is maximized for
$b\gg a,$  yielding
\begin{equation}\label{rate-cylinder}
\frac{dN}{dt} = \frac{1}{16} \frac{L \omega_0^2 \,\delta\rho_0^2}{c\, a^2}.
\end{equation}
By following the same method, one derives  
 the emission rate for parallel plates, as given by Eq.~(\ref{two-plates}), 
from the common value for the matrix element given by Eq.~(\ref{plates2}), thus 
resulting in  $1/a^2$ dependence for 
both configurations~\cite{foot}.  
From this point, the different powers of $\omega_0$ in Eq.~(\ref{two-plates}) and 
 Eq.~(\ref{rate-cylinder})
may be inferred by dimensional analysis, after taking into account the necessary factors of length and 
area for the two setups. 

When taking the same parameters discussed in connection with Eq.~(\ref{order1}), and with 
$L\omega_0/c=2\pi$ [$L= 3{\rm cm}],$ we find 
\begin{equation}
\frac{dN}{dt} = 5.6 \times 10^3 \,{\rm sec}^{-1}.
\end{equation}
As for Eq.~(\ref{order1}), this is only an estimate of the order of magnitude, since  this
value of $L$  is of the order of the typical wavelengths. In fact,  diffraction at the borders of the 
waveguide, not taken into account here, are negligible only if $L\omega_0/(2\pi c)\gg 1,$ corresponding 
to unrealistic values of $L.$  

In summary, the emission rate for a long cylindrical waveguide was computed from Hamiltonian 
first-order perturbation theory. This rate is slightly smaller than the value for parallel plates.
However, 
all photons propagate along the axis of the waveguide, whereas for the parallel plates 
the total emission rate is divided between 
all directions parallel to the plates. 
In spite of the lower orders of magnitude, 
 the dynamical Casimir effect in waveguides 
has some advantages when compared with the closed cavity setups. 
There is no need to tune the mechanical frequency $\omega_0$ into parametric resonance 
with the very sharp cavity resonances,
 and the emission rate is a simple power law function of $\omega_0.$
The generated photons are not confined in a high-Q cavity, and hence are more easily available
for detection. 
Finally, the twin photons in a given pair are created simultaneously, opening (a somewhat remote)
way for coincidence measurements~\cite{foot2}.


\begin{thebibliography}{99}
\bibitem{foot} In Ref.~\cite{two-plates}, the average photon number was computed for the following 
motion: $\delta z_1=\delta z_0 \exp(-|t|/\Delta t) \cos(\omega_0 t),$ and then  the rate was derived by
dividing it by $\Delta t.$ The result agrees with the present derivation because the integrated squared amplitude
is $\int \delta z_0^2 \exp(-2|t|/\Delta t) dt = \delta z_0 \Delta t.$ Hence the integrated effect
of the motion from $-\infty$ to $\infty$ turns out to be equivalent to the effect of a constant amplitude 
from $t=0$ to $t=\Delta t.$
\bibitem{foot2} In a closed cavity,  the leakage of the intracavity photons (and more generally 
any loss mechanism) is a random process that destroys the time correlation between the twin photons.
\end{thebibliography}
\end{document}